\documentstyle[aps,preprint,prl,epsfig]{revtex}

\begin{document}
\tightenlines
\draft

\title{Transport in finite size systems: an exit time approach}
\author{ P. Castiglione$^{(a) (b)}$, M. Cencini$^{(a)}$,
A. Vulpiani$^{(a)}$ and E. Zambianchi$^{(c)}$ }

\address{(a) Dipartimento di Fisica, Universit\`a ``La Sapienza'', and
       INFM, Unit\`a di Roma 1,\\ P.le A. Moro 2, I-00185, Roma,
       Italy} 
\address{(b) Present address: Laboratoire de Physique
       Statistique, Ecole Normale Sup\'erieure,\\ 24 rue Lhomond, 75231
       Paris, France }     
\address{(c) Istituto di Meteorologia e Oceanografia, Istituto Universitario Navale, \\Via Acton 38, I-80133 Napoli, Italy}

\date{\today}

\maketitle

\begin{abstract}

In the framework of chaotic scattering we analyze passive tracer
transport in finite systems. In particular, we study models with
open streamlines and a finite number of recirculation zones. In the
non trivial case with a small number of recirculation zones
a description by mean of asymptotic quantities (such as the eddy
diffusivity) is not appropriate. The non asymptotic properties of
dispersion are characterized by means of the exit time statistics, which
shows strong sensitivity on initial conditions. This yields a
probability distribution function with long tails, making impossible a
characterization in terms of a unique typical exit time.
\end{abstract}

\section{INTRODUCTION}
\label{sec:1}

The problem of transport in velocity fields characterized by different
flow regimes in different subareas (i.e.  a ``spatially disordered set
of streamlines'' \cite{Y88}) has been considered by many
authors. Particular attention has been devoted to steady and
time-dependent oceanic and atmospheric flows with recirculations
\cite{RY82,P88,Rich95,Buff97} and the related problem of the
dispersion in porous media \cite{Bren80,KB85,Hor96}.  In this paper we
will focus our attention on flows with recirculations of geophysical
interest.  The study of transport properties in presence of
recirculations has a crucial relevance, since gyre- or eddy-like
recirculating patterns are ubiquitous features in different areas of
the world ocean and atmosphere.  In the ocean these features are
typically induced by forcing spatial structures at the boundaries
(e.g. bottom topography, wind stress curl, coast-lines) or by intrinsic
dynamical reasons (mesoscale eddies), one interacting with the other
(for a general reference see \cite{Ol91}).

It is worth noting that large-scale meandering jets, which are
typically associated with the extensions of western boundary currents,
often separate ocean regions characterized by different
physical and biogeochemical characteristics. Consequently, the study
of mixing processes in correspondence of them is important also for
multidisciplinary investigations \cite{Bow85} (e.g.
the biological effects of longitudinal transport in western
boundary current extensions see \cite{LM94}).

The systems we consider are characterized by the joint presence of
open streamline areas where particle motion is essentially a ballistic
flight and closed streamline regions typically distributed according
to a periodic geometry, where particles tend to be trapped.  The
easiest way to study the transport properties in such systems is by
averaging over smaller space or over shorter time scales. Typically,
this results in the possibility of describing dispersion by an
equation for macroscopic quantities of the system such as the average
passive scalar concentration in terms of an effective drift and an
effective diffusivity (as classically done by Taylor \cite{Taylor53};
see \cite{Y88}, for a detailed discussion), at least in the case of
standard diffusion.  However, a description in terms of equations for
macroscopic quantities needs to average out the small scale (fast)
features of the velocity field and thus such approach applies only for
asymptotic times, when particles have been able to thoroughly sample
the different flow regimes in the system.

As stressed by Young \cite{Y88}, before reaching this asymptotic
regime very interesting transient behaviors, which cannot be described
within an effective diffusion model, could occur. The transient regime
may be very long and, if the system is finite, the asymptotic one may
not be reached. This is, typically, the case of finite domain systems
with no large scale separation between the size of the domain and the
largest characteristic Eulerian length.  In realistic flows, which are
usually characterized by a fairly limited number of recirculations,
fluid particles ordinarily sample just a fraction of the available
regimes.

Therefore, it is often not possible to characterize dispersion simply
in terms of asymptotic quantities such as average velocity and
diffusion coefficients: different approaches are needed, as done in
\cite{DUBOIS,ABCCV97,LAV98} for the characterization of transport
in closed domains; such as the symbolic dynamics approach to the
sub-diffusive behavior in a stochastic layer and to mixing in
meandering jets respectively described in \cite{Misg98} and
\cite{CLVZ99}; or the study of tracer dynamics in open flows in terms
of chaotic scattering by T\'el and coworkers
\cite{T1,T2,T3}; and the exit time approach of \cite{Buff97}.

The aim of this paper is to describe dispersion in finite size
systems.  In particular, we want to characterize flows with a small
number of recirculations using some ideas stemming from the chaotic
scattering theory.

In Section~\ref{sec:2} we briefly review some methods for the study of
transport in non asymptotic regimes, namely the
finite-size diffusion coefficients and the exit time statistics
originated from the chaotic scattering phenomenon.
Section~\ref{sec:3} contains a description of the two 
studied models (traveling wave and meandering jet) and some numerical
results.  Section~\ref{sec:4} is devoted to the comparison between
numerical results and a probabilistic model. Conclusions are presented
in Section~\ref{sec:5}.

\section{Tools for the study of non-asymptotic \\
transport properties}
\label{sec:2}

The investigation of passive tracer diffusion is usually reduced to
the study of an effective equation describing the long-time,
large-distance average tracer concentration behavior.  Under rather
general hypotheses, given an Eulerian velocity field the long-time
transport process is uniquely characterized by the effective diffusion
(or diffusivity) tensor ${D^E}_{ij}$:
\begin{equation}
{D^E}_{ij}= \lim_{t\rightarrow \infty} \frac {1}{2t}
<(x_{i}(t)-<x_{i}>)(x_{j}(t)-<x_{j}>)>\,,
\label{eq:eddydiff}
\end{equation}
where ${\bf x}(t)$ is the position of the tracer particle at time $t$;
$i,j=1,\cdots,d$, and $d$ is the spatial dimension; the average is
taken over the tracer initial conditions or, equivalently, over an
ensemble of tracer particles.  The effective diffusion tensor
$D_{ij}^{E}$ takes into account the molecular diffusivity and the
details of the velocity field. Even in presence of simple Eulerian
fields (e.g. laminar and periodic in time) the diffusion coefficient
as a function of the parameters of the velocity field can display a
rather non trivial behavior
\cite{lagran,NEW}

It is worth underlining that the diffusivity tensor
(\ref{eq:eddydiff}) is mathematically well defined only in the
asymptotic limit, therefore its use in finite size domains yields
meaningful results only if the characteristic length $l_{u}$ of the
velocity field is much smaller than the size of the domain.  If this
is not the case (e.g. in many geophysical settings or
plasma physics \cite{DUBOIS}), dispersion can
be characterized more satisfactorily using concepts and techniques
borrowed by the dynamical systems theory which will be discussed in
the following.

As an example, consider the relative diffusion of a cloud of $\cal
{N}$ test particles in a smooth velocity field with characteristic
length $l_{u}$ and assume that the Lagrangian motion is chaotic.
Denoting with $R(t)$ the radius of the cloud:
\begin{equation} 
R^{2}(t)= {1 \over {\cal N}} \sum_{k=1}^{\cal N}|{\bf
x}_{k}(t)-\langle{\bf x}(t)\rangle|^{2}\,,
\label{def:disprel}
\end{equation} 
with
\begin{equation} 
\langle{\bf x}(t)\rangle={1 \over {\cal N}} \sum_{n=1}^{\cal N} {\bf x}_n(t)
\end{equation} 
we expect the following regimes to hold
\begin{equation} 
\overline{R^{2}(t)} \simeq \left\{ 
\begin{array}{ll}
R^{2}(0)\exp(L(2)t) & \;\;\;\;
{\mbox {if    $\overline{R^{2}(t)}^{1/2} \ll l_{u}$}}
 \\
2 {D^E} t & \;\;\;\;
{\mbox {if    $\overline{R^{2}(t)}^{1/2} \gg l_{u}$}}
\end{array}
\label{eq:regimiperR}
\right.
\,,
\label{example1} 
\end{equation}
where $L(2) \geq 2\lambda$ is the generalized Lyapunov exponent
\cite{PV87}, $D^E$ is the diffusion coefficient under the assumption
of isotropy, ${D^E}_{ij} = \delta_{ij}D^E$, and the over-bar denotes
the average over initial conditions of the cloud.  Equation
(\ref{example1}) states that as long as the size of the cloud is
relatively small, chaotic behavior prevails; later on, dispersion 
displays a diffusive behavior.

Another way to describe the above behavior, even though at first
glance rather artificial, is by introducing the ``doubling time\/''
$\cal{T}(\delta)$ at scale $\delta$ as follows: define a series of
thresholds $\delta^{(n)}= r^{n} \delta^{(0)}$, where $\delta^{(0)}$ is
the initial size of the cloud, and then measure the time
$T(\delta^{(0)})$ it takes for the growth from $\delta^{(0)}$ to
$\delta^{(1)}= r \delta^{(0)}$, and so on for
$T(\delta^{(1)})\,,\;T(\delta^{(2)})\,,\ldots$ up to the largest
considered scale.  Though strictly speaking the term ``doubling time''
refers to the threshold rate $r=2$, any value can be chosen for $r$,
even if a too large one might not separate different scale
contributions.

Performing ${\cal   M}   \gg 1$ experiments    with  different initial
conditions  for  the  cloud,  we   define   an  average doubling  time
$\cal{T}(\delta)$ at scale $\delta$ as
\begin{equation} 
{\cal T}(\delta) = < T(\delta) >_{\cal M} =\frac{1}{{\cal M}}
 \sum_{m=1}^{{\cal M}} T_{m}(\delta)\,.
\label{def:taudelta}
\end{equation} 
It is worth noting that the average in (\ref{def:taudelta}) is
different from the usual time average (see \cite{ABCPV96} for a
detailed discussion of this point).  A finite size diffusion
coefficient $D(\delta)$ can then be introduced as
\begin{equation} 
D(\delta)=\delta^{2}\lambda(\delta)\,,
\label{def:fsd}
\end{equation} 
where
\begin{equation} 
\lambda(\delta)=\frac{\ln r}{\cal {T}(\delta)}
\end{equation} 
is the finite size Lagrangian Lyapunov exponent \cite{ABCCV97}.  It
can be shown that the Lyapunov exponent $\lambda$ can be obtained from
$\lambda(\delta)$ for $\delta \rightarrow 0$, namely $\lambda(\delta)
= \lambda$ for $\delta \ll l_{u}$.

For a tracer cloud of non-infinitesimal size ${\cal T}(\delta)$ 
depends on the details of the nonlinear mechanisms  of 
expansion: in the case of standard diffusion $D(\delta)$ is a constant, i.e. 
$1/{\cal T}(\delta) \sim \delta^{-2}$ \cite{ABCCV97}.
Thus
\begin{equation} 
\lambda(\delta) \simeq \left\{ 
\begin{array}{ll}
\lambda & \;\;\;\;
{\mbox {if    $\delta \ll l_{u}$}}
 \\
D/\delta^{2} & \;\;\;\;
{\mbox {if    $\delta \gg l_{u}$}}\;\;.
\end{array}
\right.
\label{eq:regimipertau} 
\end{equation} 

The fixed scale analysis allows us to extract physical information at different
spatial scales avoiding some unpleasant consequences resulting
from working at a fixed delay time $t$.  For instance, in presence of
strong intermittency, $R^{2}(t)$ as a function of $t$ can be rather
different from one realization to another.  In figure 1a we present an
example taken from \cite{ABCCV97} where different
exponential rates of growth for different realizations of $R^{2}(t)$
produce a spurious behavior of $\overline{R^{2}(t)}$.  Since at
small scales, say $l_u$, one typically has an exponential growth of
$R(t)$ due to Lagrangian chaos and at large scale a diffusive
behavior, one can mimic $R(t)$ as follows:
\begin{equation}
R^2(t)= \left\{ 
\begin{array}{ll}
\delta_0^2 e^{2 \, \gamma \, t} & R(t)< l_u\\
2 D (t-t_*) & R(t)> l_u
\end{array}
\right.
\label{eq:9}
\end{equation}
where $t_*$ is such that $R(t)$ is continuous.\\
Figure 1b 
shows the average $\overline{R^{2}(t)}$ versus time $t$; at large times
the diffusive behavior is recovered, 
but at intermediate times 
an apparently anomalous regime occurs.
This last regime is only due to the superposition of
exponential and diffusive contributions by different samples at the
same time.  On the other hand the doubling time analysis
yields unambiguous results, see figure 1c.

Let us remark that the above technique recovers the
usual asymptotic description when there is a large scales separation 
and also if a genuine anomalous diffusion occurs \cite{zav}, 
in addition, it constitutes a systematic method to treat
situations in which the scales are not well separated.  Moreover,
the finite size diffusion coefficient $D(\delta)$ enables the
understanding of the different spreading mechanisms at different
scales.  This has been recently shown in
\cite{LAV98}, where this method is applied to analyzing experimental
trajectories described by surface drifters in the Adriatic Sea.

Another interesting approach to the study of transport properties in
finite systems is the {\it chaotic scattering theory} used in
\cite{T1,T2,T3} for passive tracer advection in open flows.  In a
nutshell, chaotic scattering can be summarized as follows.  A particle
arriving from, say, $x=-\infty$ enters a region (defined as the
scattering, or interacting, region) where due to the presence of a
potential it scatters, then exits and goes to $x=\infty$.  Typically,
for a rather general class of potentials \cite{Ott} the time a
particle takes to escape from the scattering region can be very
sensitive on the impact parameter $b$ and thus displaying a {\em
chaotic} character (like a ball in a pinball game).  This justifies
the definition in terms of chaos, even if since chaos is a
time-asymptotic concept, from a technical point of view this kind of
behavior is not chaotic: particles after a transient (even if very
long) leave the scattering region and enter a regular motion regime.

A nice example of the application of chaotic scattering theory in
passive tracer study is given in
\cite{T1,T2} for the motion of Lagrangian tracers in blinking
vortex-sink system and in a von Karman vortex street behind a cylinder
in a channel. Tracer particles can temporarily be trapped in certain
regions, e.g. the wakes of the von Karman street, performing very
irregular paths. On the other hand, since the non stationarity of the
flow is mainly restricted to a finite mixing region around the
obstacles, the asymptotic almost free particle motion is also
recovered.

For the use of the exit time approach for the transport and mixing in
volume preserving maps see Ref. \cite{meiss}.

The analogy between chaotic scattering, occurring in Hamiltonian
systems, and passive scalar motion can be drawn in formal terms for
two dimensional incompressible velocity field \cite{T1}. In this case the
Eulerian field is described by a stream-function $\psi(x,y,t)$, and
the corresponding equations for the Lagrangian evolution are:
\begin{equation} 
\frac{{\rm d}x}{{\rm d}t}=-\frac{\partial \psi(x,y,t)}{\partial y}\,,\;\;
\frac{{\rm d}y}{{\rm d}t}=\frac{\partial \psi(x,y,t)}{\partial x}\,.
\label{eq:2.9}
\end{equation}
The equations (\ref{eq:2.9}) are nothing but the canonical equations
for a one dimensional time dependent Hamiltonian system, where
the stream function plays the role of the Hamiltonian.

Since in chaotic scattering the particle exit time from the
interacting region depends strongly on the initial conditions, it is
interesting to look at the time delay function \cite{T1}, i.e. the
exit time as a function of the initial position, $\tau(x(0),y(0))$,
e.g. with $x(0)=x_0$. In presence of chaotic scattering
$\tau(x_0,y(0)) $, displays a rather irregular shape (see, e.g.,
fig. 5.18 in \cite{Ott}).

As already remarked, since in chaotic scattering the irregular
character is confined both in space and time (i.e. one has the
so-called transient chaos), the Lyapunov exponent is trivially zero;
however, a sort of high sensitivity on initial conditions is suggested
by the occurrence of very different time delays for very close
deployment locations.

The presence of large excursions for the exit time has an obvious
relevance for transport processes in finite size systems. The wild
variations of $\tau(x_0 ,y(0))$ pose severe limits on the possibility 
to make prediction on the particles behavior and force us to use statistical
approaches.  This leads to introduce the probability distribution
function, $P(\tau)$, of the exit times $\tau(x_0,y(0))$.

\section{Two simple flows}
\label{sec:3}

In this paper two idealized flow models are studied; both are
reminiscent of oceanographic features.  We consider finite amplitude
Rossby waves in a channel, recently revisited in \cite{LS98} (see also
\cite{Pi91,CM93}).  In the reference frame moving with the phase speed
of the wave, the flow field shows a central open streamline region
(ballistic motion) sided by trapping recirculations (see fig.~2a);
this flow pattern is a suitable, even if simplified, prototype for
studying the effect of a finite number of trapping areas on the
longitudinal dispersion of particles.

The two-dimensional incompressible Rossby wave flow \cite{LS98}
here considered is specified by the following stream-function:
\begin{equation}
\Psi_0 (x,y)= A_{0} \sin (K_{0}x) \sin (L_{0}y)-c_{0} y
\label{eq:3.1}
\end{equation}
where $A_0$ is related to the maximum velocity in the $y-$direction,
($K_0,L_0$) is the wave vector and $c_0$ is the phase speed of the
primary wave in the $x-$direction. In (\ref{eq:3.1}) $\Psi_0$ is
expressed in the reference frame co-moving with the primary wave.

In order to reproduce the instabilities usually present in geophysical
flows, we introduce a time-periodic perturbation, $\delta
\Psi(x,y,t)$:
\begin{equation}
\delta \Psi(x,y,t)= \alpha \sin ( K_{1} x-\Omega t) \sin (L_{1} y)
\label{eq:3.2}
\end{equation}
where $\alpha >0$ is a (not necessarily small) parameter which
controls the amplitude of the perturbation and ($K_1,L_1$) is the
wave-vector of the perturbation (secondary wave).  Even though
realistic disturbances cannot be characterized in terms of a single
wave alone \cite{wiggins}, the perturbation (\ref{eq:3.2}) is a first
step towards a description of the complex structure of transport
mechanisms in these systems.

The second flow we investigate is a meandering jet, which represents a
natural extension of the above Rossby wave system, and it was
extensively studied in the literature with particular reference to the
Gulf Stream \cite{Bow85,Watts83}.  Again, the flow can be subdivided
into different regions roughly corresponding to a prograde flow (in
reality, the current jet core; in our schematization, the open
streamline regime), recirculation regions and, at a farther
distance, an essentially quiescent (far) field (see fig.~2b).

We consider now fluid particle trajectories in the two-dimensional
kinematic model originally proposed by Bower \cite{B91} and thereafter
widely studied \cite{CLVZ99,Sam92,griffa,Cu93}.  The large-scale flow,
in a reference frame moving eastward with a velocity coinciding with
the meander phase speed and suitably nondimensionalized, is expressed
by the stream function:
\begin{equation}
\psi(x,y) \,=\,-\tanh\left[ \frac{y-B\cos\, kx}{(1+k^2B^2\sin^2kx)^{1/2}} \right] 
+ cy\,.
\label{eq:3.2.1}
\end{equation}
In fig.~2b we show the streamlines in the co-moving
frame.

As mentioned above, chaotic advection may be introduced trough a time
dependence.  Among the different mechanisms discussed in \cite{Sam92},
we chose here a time-periodic oscillation of the meander amplitude:
\begin{equation} B(t) \,=\, B_0 + \gamma
\cdot \cos(\omega t + \theta)\,,
\label{eq:3.2.2}
\end{equation}
where we set $B_0 = 1.2$, $\gamma = 0.3$, $\omega = 0.4$ and $\theta =
\pi/2$. The parameters choice is mainly motivated by observations and
numerical results, as discussed in \cite{CLVZ99}.

It is worth underlining that, even if quite a great deal of efforts
has been devoted to study fluid exchange across the jet
\cite{B91,Sam92,griffa}, very little is known as to tracer behavior in
the along-jet direction (periodic flows with open streamlines have
been proposed as model for the meandering jets; the presence of
recirculations has been seen to induce non-trivial effects on the
along-jet dispersion, see \cite{LPVZ96}).

Tracer particle trajectories have been numerically generated from
eqs. (\ref{eq:2.9}) with the stream-functions corresponding to the
traveling wave (\ref{eq:3.1}-\ref{eq:3.2}) and the meandering jet
(\ref{eq:3.2.1}-\ref{eq:3.2.2}).  However, since the results relative
to the two flows are qualitatively the same, we shall present and discuss
just those obtained for the meandering jet.  

As can be seen from the streamfunctions, the two flows are periodic in
the longitudinal direction. Since we are interested in the
characterization of longitudinal transport, the number of elementary
flow structures (or cells), $N_c$, constituting the system, is a
crucial parameter.  For very large $N_c$, the dispersion properties of
the system can be studied using asymptotic techniques, e.g. the
multiscale method \cite{LPP,BLP}. On the contrary, we mainly
concentrate on systems with a small number of cells, namely $N_c
\simeq 2-10$.

The first focus of our analysis is the particle exit time (or time
delay function, see Sect.~\ref{sec:2}) as a function of the initial
position, i.e. the time $\tau(x_0,y(0))$ a tracer particle deployed at
$(x_0,y(0))$ takes to reach the boundary $x_{max}=N_c\,2 \pi/k$.  Two
very different scenarios occur for large and small $N_c$.

Figs.~3a-d show the behavior of $\tau(x_0 ,y(0))$ for
$N_c=3,10,100,1000$.  Increasing the system size there is a clear change 
in the shape of $\tau(x_0 ,y(0))$: highly inhomogeneous structures
(fractal objects) for low  $N_c$  (figs.~3a,b),
with very strong fluctuations of the exit time value even for small
variations of the initial conditions.  As $N_c$ increases, the shape
of $\tau(x_0 ,y(0))$ becomes more and more homogeneous.

The fractal character of $\tau(x_0 ,y(0))$ is evident from figs.~3a
and the enlargements figs.~4a-b, which suggest the self-similarity of
the structures at different scales. This can be quantitatively
assessed studying the correlation dimension $\cal D$ of the initial
condition set $\{ y(0)\}$ such that $\tau(x_0,y(0))>\Theta$. Using the
Grassberger and Procaccia algorithm \cite{GP}, i.e.  computing the
percentage $C(r)$ of pairs $(y_i,y_j)$ such that $\mid y_i-y_j \mid
\leq r$; for small $r$ we obtain $C(r) \sim r^{\cal D}$ (shown in
fig.~5) with ${\cal D}<1$. The value of ${\cal D}$ can depend weakly
on the threshold $\Theta$, e.g. for $\Theta=15$ and the parameters of
fig.~3a ${\cal D}$ results $0.83$ and $\Theta=30$ yields ${\cal
D}=0.78$

Let us remark that also for very large values of $N_c$ some fractal
structures may be present, in particular, on very small scales
(detectable only for infinitesimally close particles). However, we do
not consider this feature because on those scales in real fluids we
expect the presence of smoothing due to molecular diffusion.

In order to characterize the system behavior, we have also studied
the probability density function $P_{N_c}(\tau)$.  In figs.~6a-d the
probability density functions $P_{N_c}(\tau)$ corresponding to
$N_c=3,10,100,1000$ are shown.  For large $N_c$ (fig.~6c,d)
$P_{N_c}(\tau)$ displays an asymptotic shape which can be obtained
with simple probabilistic arguments (see below, Sect.~\ref{sec:4}),
whereas in the opposite case (fig.~6a,b) the distributions exhibit
sharp peaks in correspondence of the ballistic time and exponential
tails indicating the possibility of very large excursions.

The above results show that the dispersion process in a
finite size system cannot be described in terms of a unique
characteristic time. As shown in figs.~3a,b, for $N_c=3-10$ the exit
time $\tau(x_0,y(0))$ exhibits very strong fluctuations within $3\!-\!4$
orders of magnitude, which prevents the possibility to extract
meaningful information just from the average of $\tau(x_0,y(0))$.
This is reflected in the exit times probability distribution: for large
$N_c$ the shape of the distribution suggests the appropriateness of
the average exit time to describe the diffusion, whereas this is not
the case for small $N_c$, where $P_{N_c}(\tau)$ exhibits exponential tails.

\section{A probabilistic model \\compared with numerical results}
\label{sec:4}
In the previous Section we have discussed some statistical properties
of the Lagrangian dynamics generated by the streamfunctions
(\ref{eq:3.1},\ref{eq:3.2.1}); it seems thus natural to look for
probabilistic models reproducing the above properties.

The Rossby wave flow (fig.~2a)  shows two different regions; the
central one characterized by ballistic motion (open streamlines) and
particle trapping recirculations (closed streamlines) on both
sides of the ballistic regime.  The meandering jet flow (fig.~2b), in
addition to the recirculation and ballistic ones, presents two far
field regions moving retrogradely with respect to the jet core.
However, with our parameter choice the far field is practically never
visited by the tracer particles \cite{CLVZ99,Sam92} and thus we
shall not consider them.

This intrinsic subdivision of the flow fields suggests to
build a discrete symbolic picture of the motion: one can 
label the trapping and ballistic regions with the symbols $T$ and $B$  
respectively and then construct a probabilistic model.  Since the
studied flows are time periodic we express the time in number periods
of the perturbation, namely $2 \pi/\omega$.

The simplest conceivable probabilistic model is a Bernoulli scheme
such that at each time the particle can be in the states $B$ with
probability $p$ or $T$ with probability $1-p$.  However, we do not
expect such a model to give a good description of the system,
because once the length $L$ of the system (i.e. the equivalent of
$N_c$) is fixed, the probability $P_L(\tau)$ that at time $\tau \geq
L$ the particle exits from the domain depends just on the parameter
$p$, and just one free parameter is obviously not enough to
characterize the system behavior.

The natural choice for a model which take into account memory
effects is a Markov chain \cite{feller}. The process is completely
defined by the transition matrix $W_{ij}$, i.e. by the probability to
go in one step to the state $j$ starting from the state $i$
($i,j=B,T$) which has the properties:
\begin{equation}
W_{ij} \geq 0 \;\;\;\;{\rm and}\;\;\;\; \sum_{j}W_{ij}=1\;\;.
\label{ranmatr}
\end{equation} 
The probabilities $P_i$ to be in state $i$ are
given by
\begin{equation}
P_i=\sum_j P_j \, W_{ji}.
\label{invar}
\end{equation}

Now we consider the following stochastic process for the evolution of a 
particle
\begin{equation}
\Delta x(t)= \left \{ 
\begin{array} {ll} 
1& \:\: {\mbox {with probability}}\:\: W_{iB} \\
0& \:\: {\mbox {with probability}}\:\: W_{iT} \,,
\end{array}  
\right.
\label{markov}
\end{equation}
where $i$ represents the state visited at time $t-1$ and $\Delta x(t)$ the
increment in the position of the particle at time $t$.

Then we can compute the probability $P_L(\tau)$ as follows:
\begin{equation}
P_L(\tau)=\sum_{k=L-1}^{\tau-1}\,P_{L-1}(k)\,F_{BB}(\tau-k)
\label{ptl}
\end{equation}
where $F_{BB}(n)$ is the probability of first arrival from 
state $B$ to the state $B$ in $n$ steps, which obeys the following
recursive formula \cite{feller}:
\begin{equation}
F_{BB}(n)=(W^n)_{BB}-\sum_{k=1}^{n-1}\,F_{BB}(n-k)\,(W^k)_{BB}\,,
\label{ricorsF}
\end{equation}
where $W^k$ indicates the $k$-th power of the matrix $W$.
Applying recursively the (\ref{ptl}) yields
\begin{equation}
P_L(\tau)=\sum_{k_1=L-1}^{\tau-1}\,F_{BB}(\tau-k_1)\sum_{k_2=L-2}^{k_1-1}\,
F_{BB}(k_1-k_2)...\sum_{k_{L_1}=1}^{k_{L-2}-1}\,F_{BB}(k_{L-2}-k_{L-1})\,
P_1(k_{L-1})
\label{sempli}
\end{equation} 
being
\begin{equation} 
P_1(k)=\left\{ 
             \begin{array}{cc}
                P_B  & k=1\\
                P_T\,(W_{TT})^{k-2}\,W_{TB} & k\geq 2.
             \end{array}
\right.
\end{equation}

Even if, in general a one-step Markov process is not enough for a
detailed description of statistical properties of the dynamical system
\cite{CLVZ99,N1,N2}, it may result a
good description as long as $\tau$ is not too small: one indeed
expects that for large exit times the memory effects are less
important.  The expression (\ref{sempli}) for $P_L(\tau)$ has been
compared with the exit time probabilities numerically computed in
Sect.~\ref{sec:3}.  In order to carry out the comparison between the
probabilistic model and the numerical results we have first to
evaluate the parameters of the model, i.e. $L$, $P_i$, $W_{i,j}$.

In the numerical evaluation of the matrix $W_{ij}$ and the
probabilities $P_i$ we have proceeded according to the following
scheme.  Expressing the time in number of periods of the perturbation,
the transition probabilities are computed from a long
trajectory ${\bf x}_{0},{\bf x}_{1},\cdots,{\bf x}_{n}$ ($n\gg 1$) in
an infinite system ($N_c=\infty$) as
\begin{equation}
W_{ij}=\lim_{n\rightarrow \infty} \frac{N_{n}(i,j)}{N_{n}(i)}
\label{eq:matrix}
\end{equation}
where $N_{n}(i)$ is the number of times that, along the trajectory,
the particle visits the state $i$ ($i=T$ or $B$) and $N_{n}(i,j)$ is the
number of times that ${\bf x}_{t}$ is in state $i$ and
${\bf x}_{t+1}$ is in state $j$
($i,j=T,B$).  The visit probabilities $P_i$ are simply given by $\lim_{n \to
\infty} N_{n}(i)/n$.  The identification of the visited state is
performed by controlling the value of the stream function and the sign
of the velocity along the $x,y$-axis \cite{CLVZ99}.

In order to evaluate $L$ we have to take into account that in the
physical system the spatial increment {\it per} time step (i.e. during one
period of the perturbation), $\Delta x$ may vary with time.  We have
then computed the most probable value $\Delta \tilde{x}$, of
$x_{t+(2\pi/\omega)}-x_{t}$, hence we have rescaled $L$ using
$[L/\Delta \tilde{x}]=\tilde{L}$, the square brackets $[\,\cdot\,]$
here indicate the integer part of the argument.

In figs.~6a-b the probability density distributions $P_L(\tau)$ are
compared with those obtained using (\ref{sempli}).  The small shift of
the peak at the beginning of the distributions could depend on the
estimate of $\Delta \tilde{x}$.  As can be seen from the figures~6a-b, even
for $L \sim 3-10$ we obtain a good description of $P_L(\tau)$ at least
for $\tau \gg 1$, i.e. for those particles that experience a large
number of transitions between different states. Increasing $L$,
$P_L{(\tau)}$ is better and better approximated by the Markov chain
prediction. It is worth remarking that we are comparing the
numerical results with the prediction of the probabilistic model
without performing any data fitting.

On space/time scales much larger than the typical time/ space
scales of the velocity field the evolution of a test particle follows
a diffusive scenario. Thus we expect the following process to be a
good model for our systems at least in the limit $L \gg 1$ and $\tau
\gg 1$: consider the diffusive stochastic process
\begin{equation}
x(t)-x(0)=\overline{v} \,t+\sqrt{2 D}\, w(t)
\label{Langevin}
\end{equation}
where 
\begin{equation}
\overline{v}=\lim_{t \to \infty} \left< {(x(t)-x(0)) \over t} \right>
\;\;\;\; {\rm and} \;\;\;\;
D=\lim_{t \to \infty} {1 \over 2t}<{(x(t)-x(0)-\overline{v}t)}^2 >
\label{VeD}
\end{equation}
$w(t)$ is a Wiener process, i.e. a Gaussian process with 
$w(0)=0$ and
\begin{equation}
\langle w(t)\rangle=0 \;,\;\;\; \langle w(t)\,w(t')\rangle={\rm min}[t,t'].
\label{Wiener}
\end{equation}
Defining $\tau$ the first exit time i.e. the maximum time for which
\begin{equation}
x(\tau)\leq x_{max}=\overline{v} \,\tau+\sqrt{2 D}\, w(\tau)
\label{Tburger}
\end{equation} 
the probability density of $\tau$ can be calculated \cite{Burgers} obtaining
\begin{equation}
P_{x_{max}}(\tau)=\frac{\mid x_{max}\mid}{\sqrt{4 \pi\,  D \,\tau^3}} \, 
\exp \left[ -\frac{(\overline{v} \tau-x_{max} )^2}{4 \,D \,\tau}\right].
\label{PTburger}
\end{equation}  
The maximum of $P_{x_{max}}(\tau)$ is reached for $\tau_{max}$ which can be estimated,
if $D/\overline{v}$ is not too large, as
\begin{equation}
\tau_{max}\simeq \frac{x_{max}}{\overline{v}}\,.
\label{Tmax}
\end{equation}

The quantities $\overline{v}$ and $D$ can be evaluated in terms of the
exit time statistics as follows
\begin{equation}
\overline{v}=\frac{L}{<\tau>_e} \;\;\;\; 
D=\frac{\left<\left( L- \overline{v} \tau \right)^2 
\right>_e} {2 \;<\tau>_e}
\label{vDmedi}
\end{equation}
where the ${\langle \;\cdot\; \rangle}_e$ indicates the average over 
the ensemble of $N_p$ particles, i.e. 
${\langle f \rangle}_e=(1/N_p)\,\sum_{i=1}^{N_p} f_{i}$.

In figs.~6a-d the probability distributions calculated in
Section~\ref{sec:3} are compared, at different values of
$L=x_{max}$, with the results given by (\ref{PTburger}) with
$\overline{v}$ and $D$ obtained by (\ref{vDmedi}) for large $N_c$.

The larger $L$, the better is the fitting with numerical data,
nevertheless the probability density function (\ref{PTburger}) does not
capture the tail behavior of the physical system probability
$P_L(\tau)$.  This is not very surprising because the process described by
eq.(\ref{Langevin}) is $\delta$-correlated in time 
and thus it is not able to describe long-range correlated
events \cite{CLVZ99} responsible of such a tail. Moreover, at
variance with the physical system, in model (\ref{Langevin}) the
velocity does not have any bound.

\section{Conclusion and discussions}
\label{sec:5}

In this paper we have studied non asymptotic properties for passive
tracer transport. For a finite system (of size $L$),
which is a rather common case in real problems (e.g. geophysical and
plasma flows), the usual characterization by means of asymptotic
quantities (such as the  eddy diffusivity) is not appropriate
if $L$ is not very large with respect to the typical length scale of the
velocity field.
  
We have considered two models of geophysical interest (traveling
waves and meandering jet) studying their Lagrangian transport
properties at varying the longitudinal size $L$. Transport has been
analyzed in terms of the statistical properties of the exit times of
particles from the systems.  This analysis has been carried out
borrowing concepts from chaotic scattering theory.

In the limit of very large $L$ the usual asymptotic scenario is recovered,
i.e. the mean velocity $\overline{v}$ and  the diffusion coefficient 
$D$ completely characterize the transport process. In this case, 
one typical time is enough to describe the basic features
of the process. On the contrary, in the more interesting (and realistic) 
cases with a not very large $L$, there is no unique characteristic 
time.  Indeed the exit times show a strong sensitivity to initial 
conditions (this is a manifestation of transient chaos in a 
non-chaotic system), limiting the possibility of a detailed forecasting
of particle behavior. This limitation is mirrored in the probability 
distribution function 
of the exit times which displays rather long tails, making impossible the 
characterization in terms of a unique time (i.e. the average exit time).

Suitable processes (e.g. Markov chains) prove to capture the relevant
statistical aspects of transport process.  If $L$ is very large the
exit times statistics is well described in terms of first exit
time problem for a linear Langevin equation involving only ${\overline
v}$ and $D$. For systems with a moderate number of recirculation zones
one has to introduce a more detailed probabilistic model.

\section{Acknowledgements}
We thank M. Falcioni, G. Lacorata and P. Muratore Ginanneschi for useful
suggestions and discussions. A particular acknowledgement to B. Marani
for the continuous and warm encouragement.  We are grateful to the
ESF-TAO ({\it Transport Processes in the Atmosphere and the Oceans})
Scientific Programme for providing meeting opportunities.  This paper
has been partly supported by INFM (Progetto di Ricerca Avanzato
PRA-TURBO), CNR, MURST (no. 9702265437), and the European Network {\it
Intermittency in Turbulent Systems} contract number FMRX-CT98-0175.


\newpage
\centerline {\bf TABLES}
\vspace{1cm}
Table 1

Transition matrix elements
\vspace{.5cm}

\begin{tabular}{|l|l|}\hline 
 $W_{BB}$ & .66 \\ \hline
 $W_{BT}$ & .34 \\ \hline  
 $W_{TB}$ & .12 \\ \hline  
 $W_{TT}$ & .88 \\ \hline  
\end{tabular}

\vspace{1cm}
Table 2

Visit probabilities
\vspace{.5cm}

\begin{tabular}{|l|l|}\hline 
 $P_{B}$ & .26\\ \hline
 $P_{T}$ & .74\\ \hline  
\end{tabular}

\vspace{1cm}

The probabilities are evaluated for the stream  function (\ref{eq:3.2.1})
with parameters $k=4\pi/15,\;B_0=1.2,\;c=0.12,\;\omega=0.4$ and
$\gamma=0.3$.

The statistics have been computed over $2\,10^{6}$ periods.

\newpage

\centerline{\bf FIGURE CAPTIONS}
\begin{description}
\item [FIGURE 1]:(a) Three realizations of $R^2(t)$ built as
eq. (\ref{eq:9}) with: $\gamma=0.08, 0.05, 0.3$ and $\delta_0=10^{-7},
\,\; D=1.5$.\\ (b) $\overline{R^2(t)}$ as function of $t$ averaged on
the three realizations shown in figure 1a. The apparent anomalous
regime and the diffusive one are shown.\\ (c) $\lambda(\delta)$ vs
$\delta$, with Lyapunov and diffusive regimes.

\item[FIGURE 2]:(a) Streamlines for the time-independent Rossby wave
flow with: $A_0=L_0=K_0=1$, $c_0=0.5$, $\alpha=0$.  
The labels $T$ and $B$ refer to the trapping regions
(recirculations) and ballistic ones (the jet channel)\\ (b) The same
as figure (a) for the meandering jet flow with:
$k=4\pi/15,\;B_0=1.2,\;c=0.12$ and $\gamma=0$.

\item[FIGURE 3]$\tau(x_0,y(0))$ for the meandering jet flow with
parameters for the unperturbed streamfunction as fig.2b and 
$\omega=0.4\,,\;\gamma=0.3\,\;\theta = \pi/2$, and  $x_0=0.1$,
$y(0) \in[-3,3]$ for (a) $N_c=3$ cells,(b) $N_c=10$, (c) $N_c=100$ and
(d) (a) $N_c=100$. 

\item[FIGURE 4] Two enlargements of figure 3a: (a) for $y(0) \in [1.15:1.20]$
and (b) $y(0) \in [1.1910:1.1925]$.

\item[FIGURE 5]Correlation integral $C(r)$ versus $r$ computed
from the data of fig. 3a,  with threshold $\Theta=15$. The straight line
has slope ${\cal D}=0.83$.

\item[FIGURE 6]$P_{N_c}(\tau)$ versus $\tau$ for the meandering jet
flow with the same parameters as in fig.3, computed with $4\,10^4$
particles starting in $x(0) \in[0,\pi/k]$, $y(0) \in[-3,3]$ for (a)
$N_c=3$ cells,(b) $N_c=10$, (c) $N_c=100$ and (d) $N_c=1000$.  The
prediction (\ref{PTburger}) is shown as dashed continuous line in all
the cases.  The parameters $\overline{v}$ and $D$ have been calculated
from (\ref{vDmedi}) for $N_c=1000$ and evaluated as $0.365$ and
$1.038$ respectively.  In figure (a) and (b) the Markovian prediction
is also shown as continuous lines using $\Delta \tilde{x}=1.36$ and
evaluating the probabilities $P_{i}$ and $W_{ij}$ as reported in tables
$1$ and $2$.

\end{description}
\newpage
\begin{figure}[p]
\centerline{\epsfig{figure=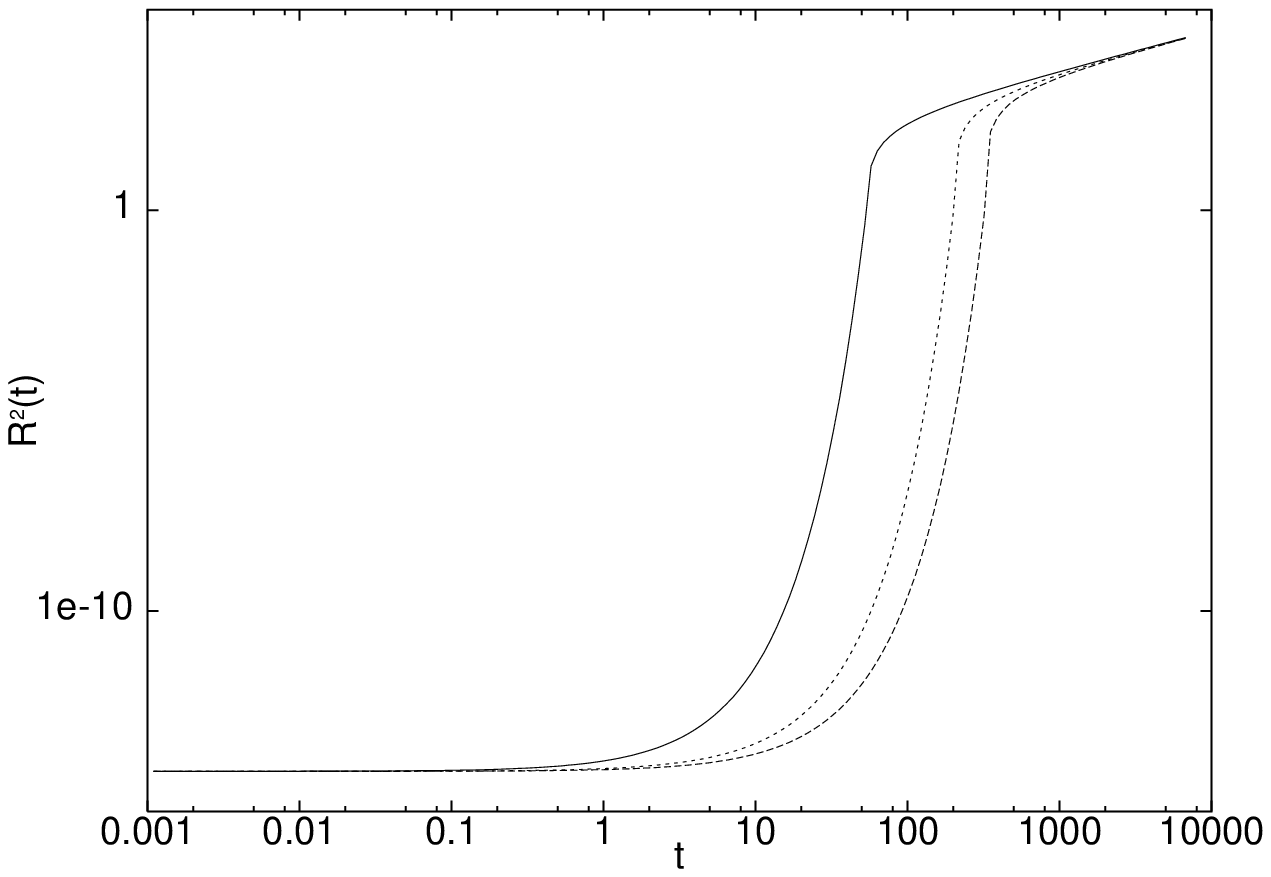,width=7cm,angle=0}}
\vspace{.2cm}
\centerline{\hspace{.5cm} (a)}
\vspace{.8cm}
\centerline{\epsfig{figure=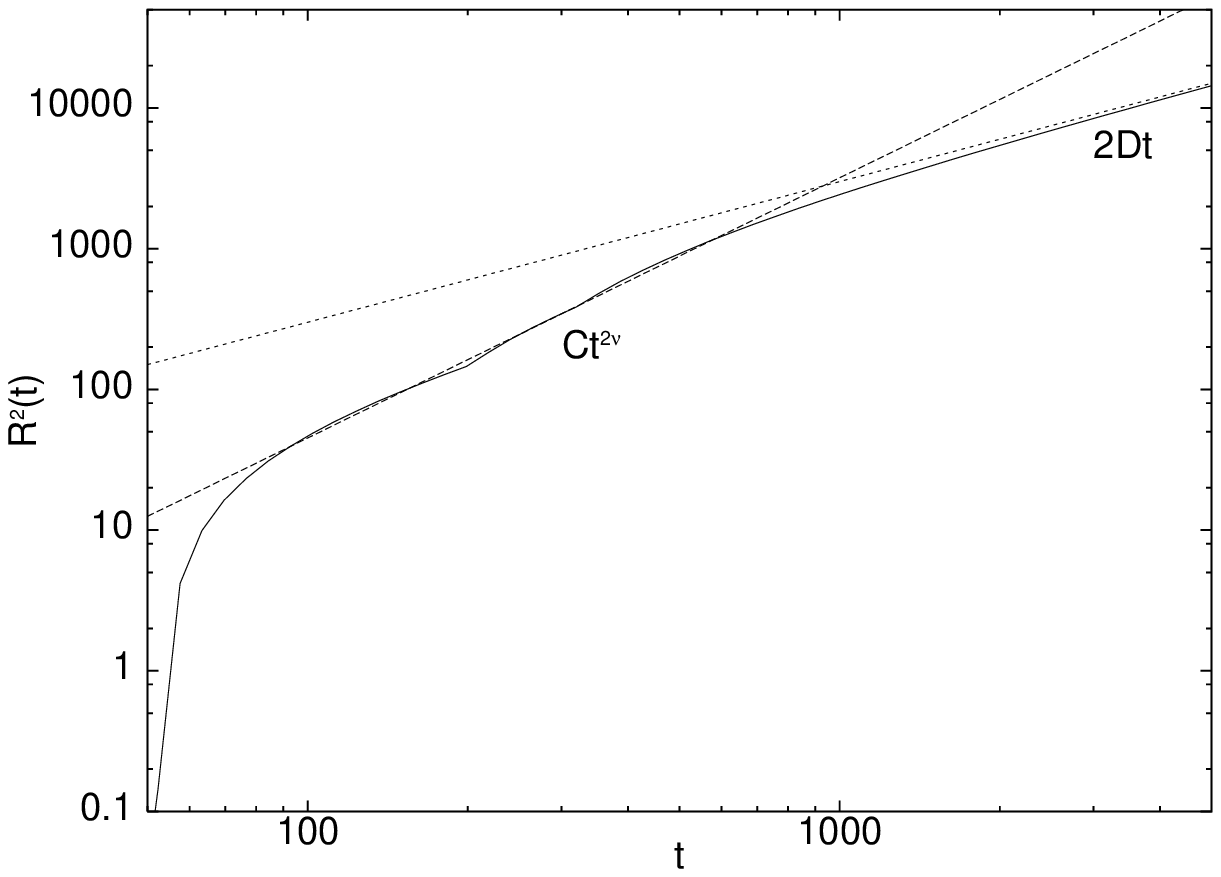,width=7cm,angle=0}}
\vspace{.2cm}
\centerline{\hspace{.5cm} (b)}
\vspace{.8cm}
\centerline{\epsfig{figure=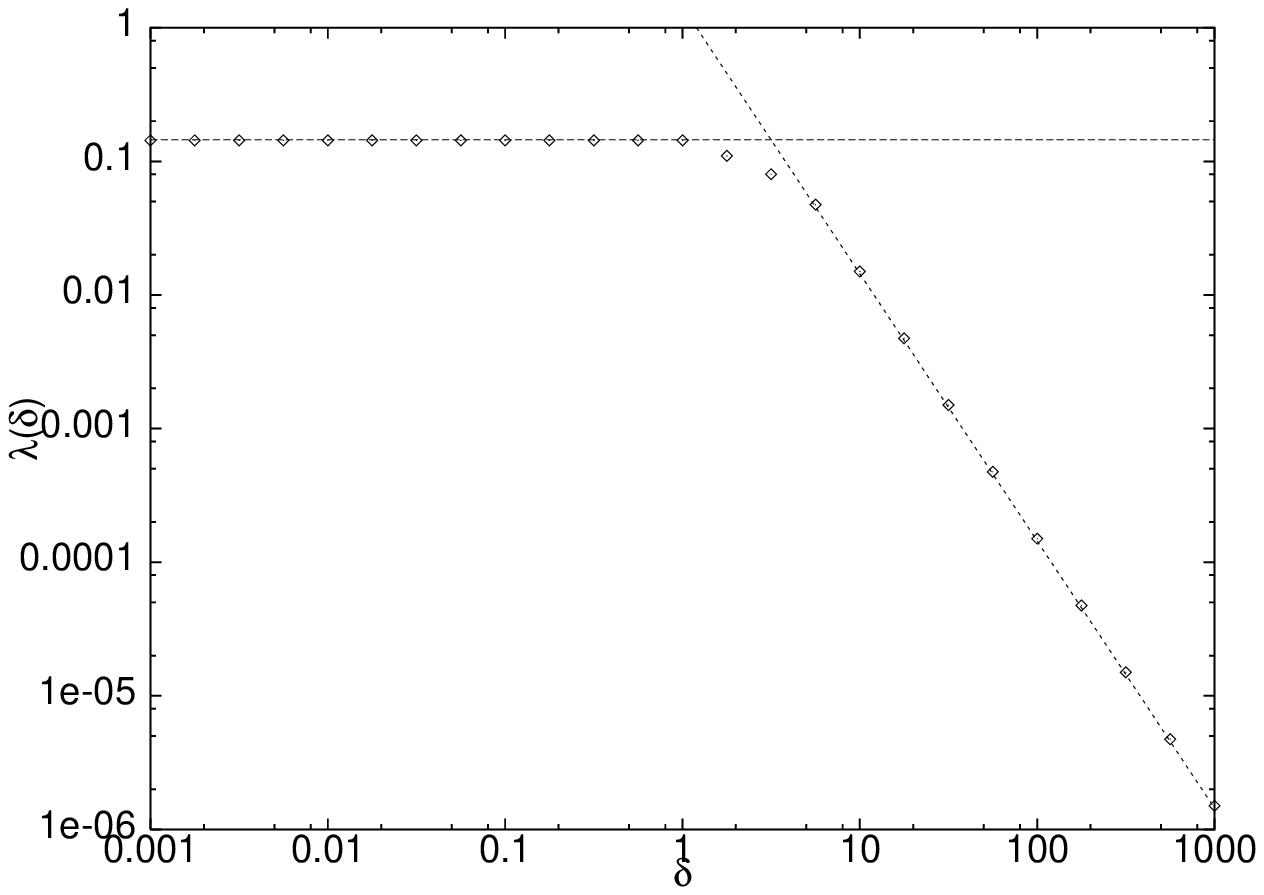,width=7cm,angle=0}}
\vspace{.2cm}
\centerline{\hspace{.5cm} (c)}
\caption{
}  
\label{fig:1}
\end{figure}

\newpage

\begin{figure}[p]
\centerline{\epsfig{figure=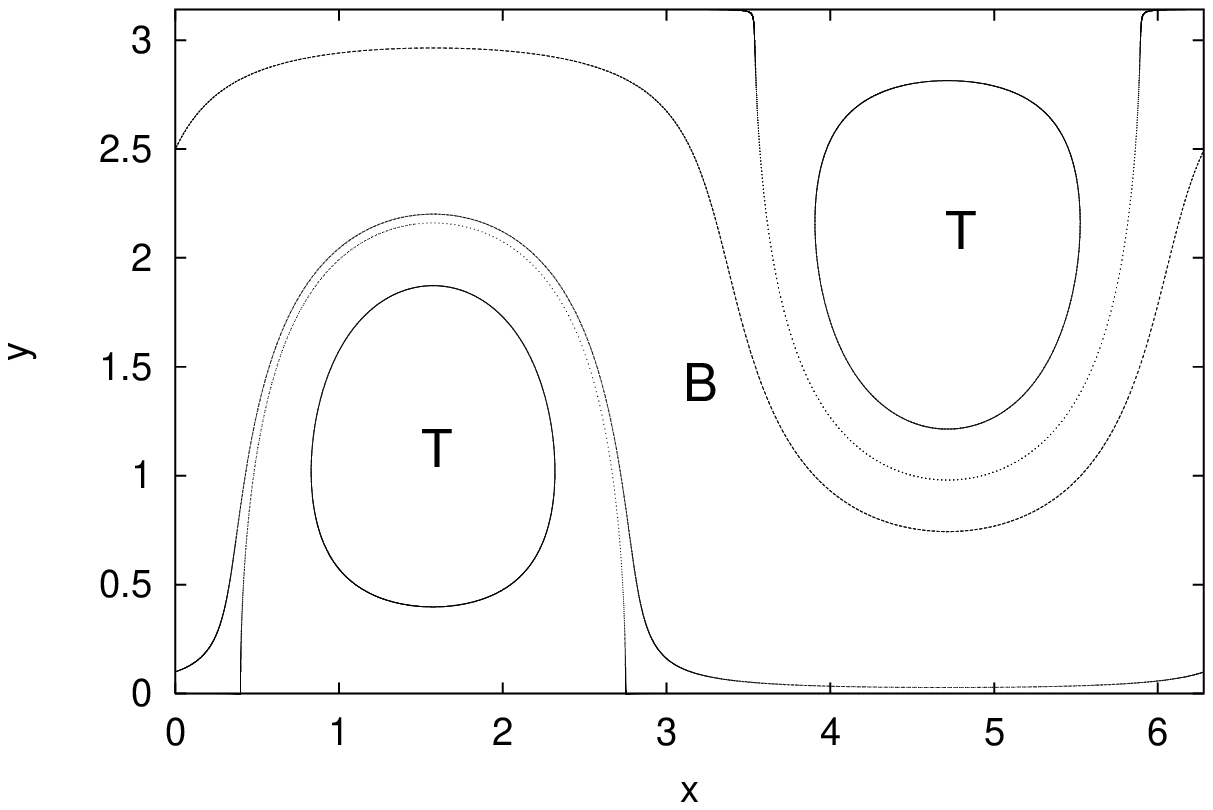,width=9cm,angle=0}}
\centerline{\hspace{1cm} (a)}
\vspace{2.0cm}
\centerline{\epsfig{figure=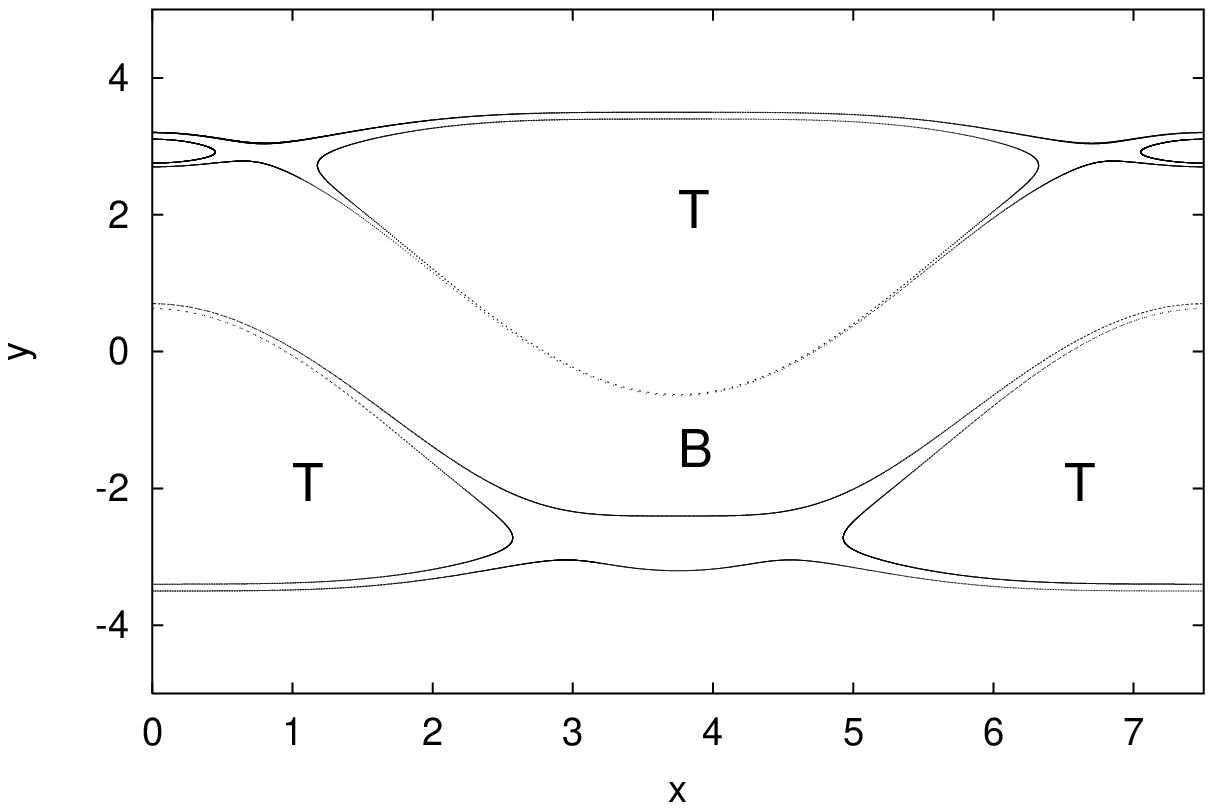,width=9cm,angle=0}} 
\centerline{\hspace{1cm} (b)}
\caption{ }
\label{fig:2}
\end{figure}

\newpage

\begin{figure}[p]
\centerline{\hspace{1cm}(a) \hspace{7.4cm}  (b)}
\centerline{\epsfig{figure=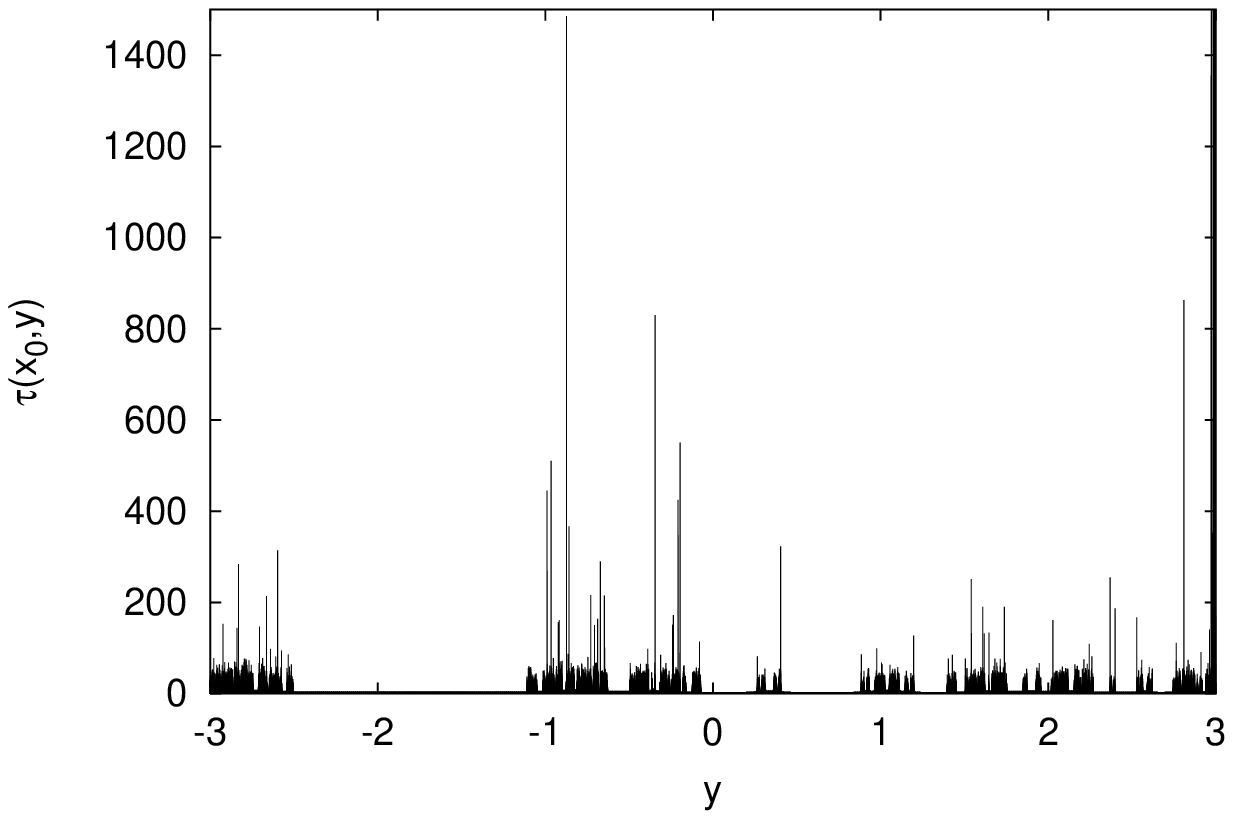,width=8cm,angle=0} \hfill 
\epsfig{figure=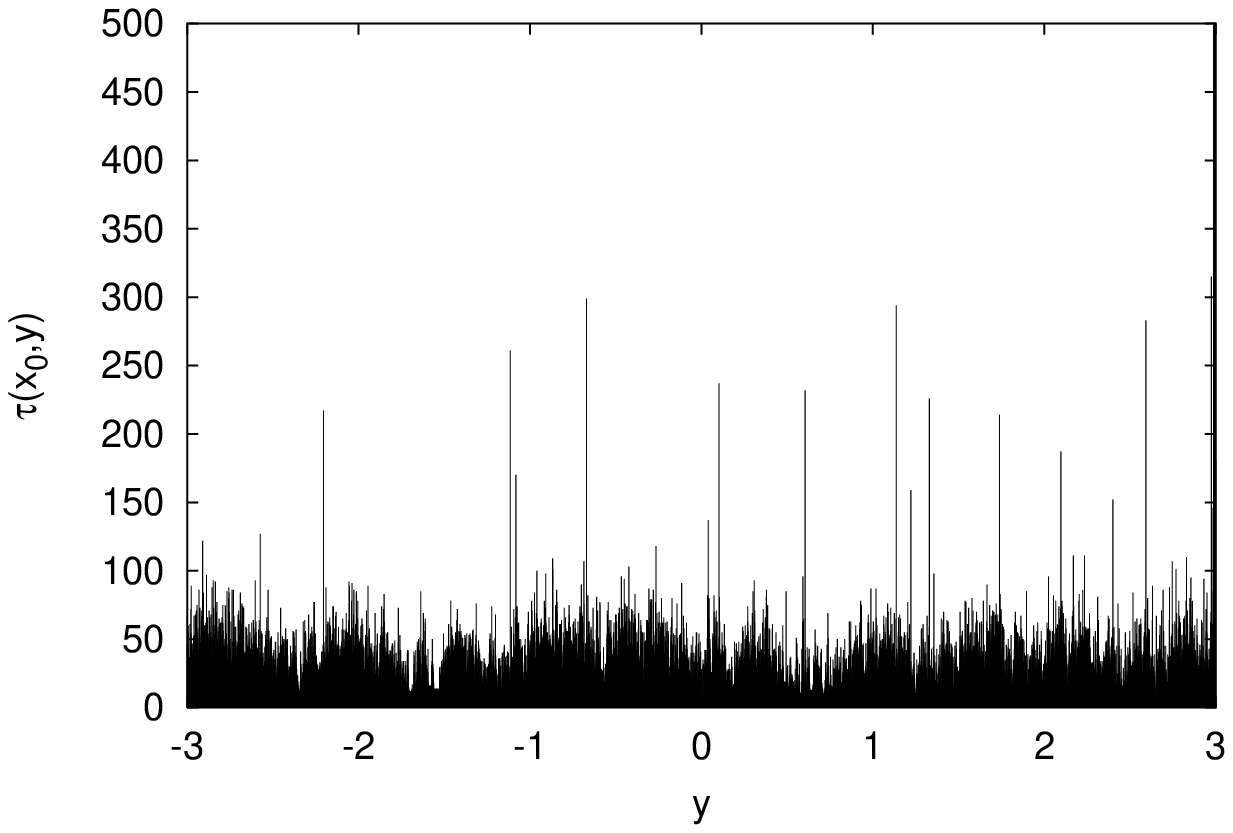,width=8cm,angle=0} }
\vspace{1.cm}
\centerline{\hspace{1cm}(c) \hspace{7.4cm}  (d)}
\centerline{\epsfig{figure=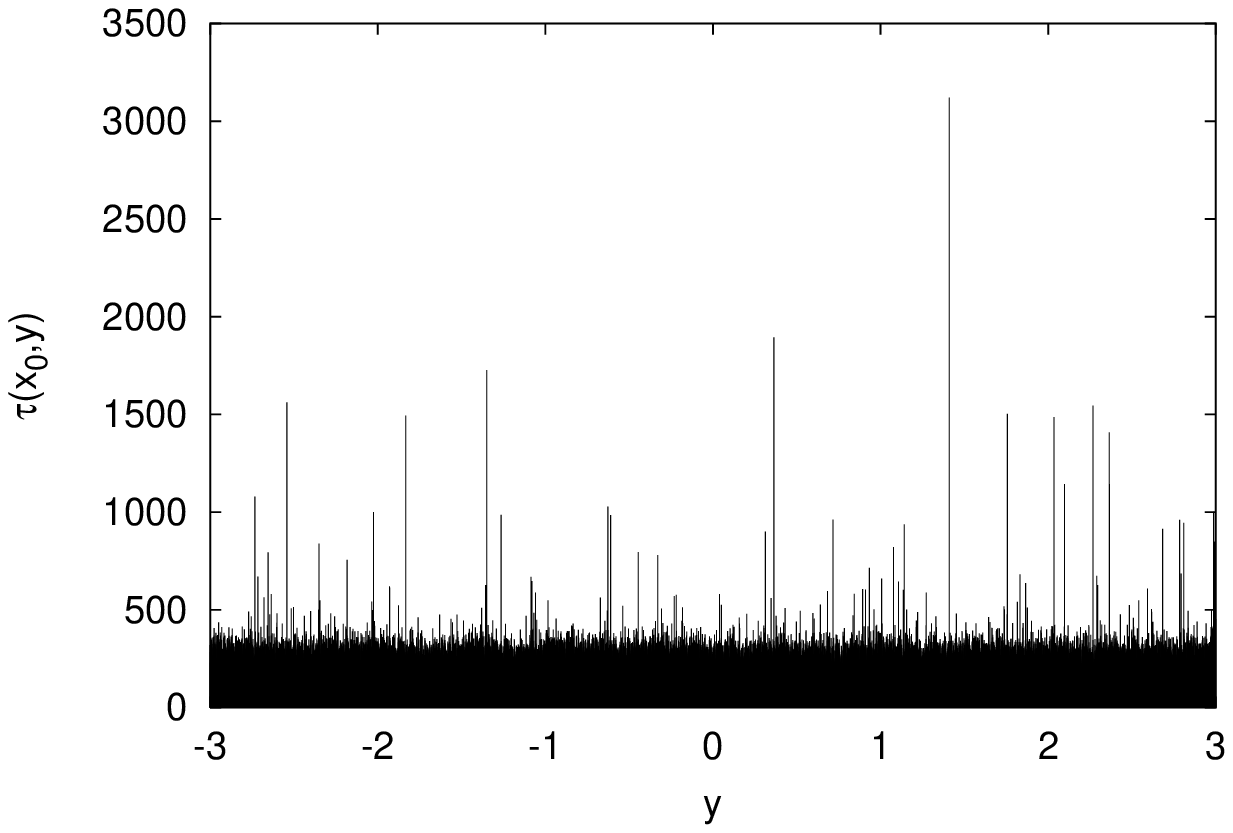,width=8cm,angle=0} \hfill 
\epsfig{figure=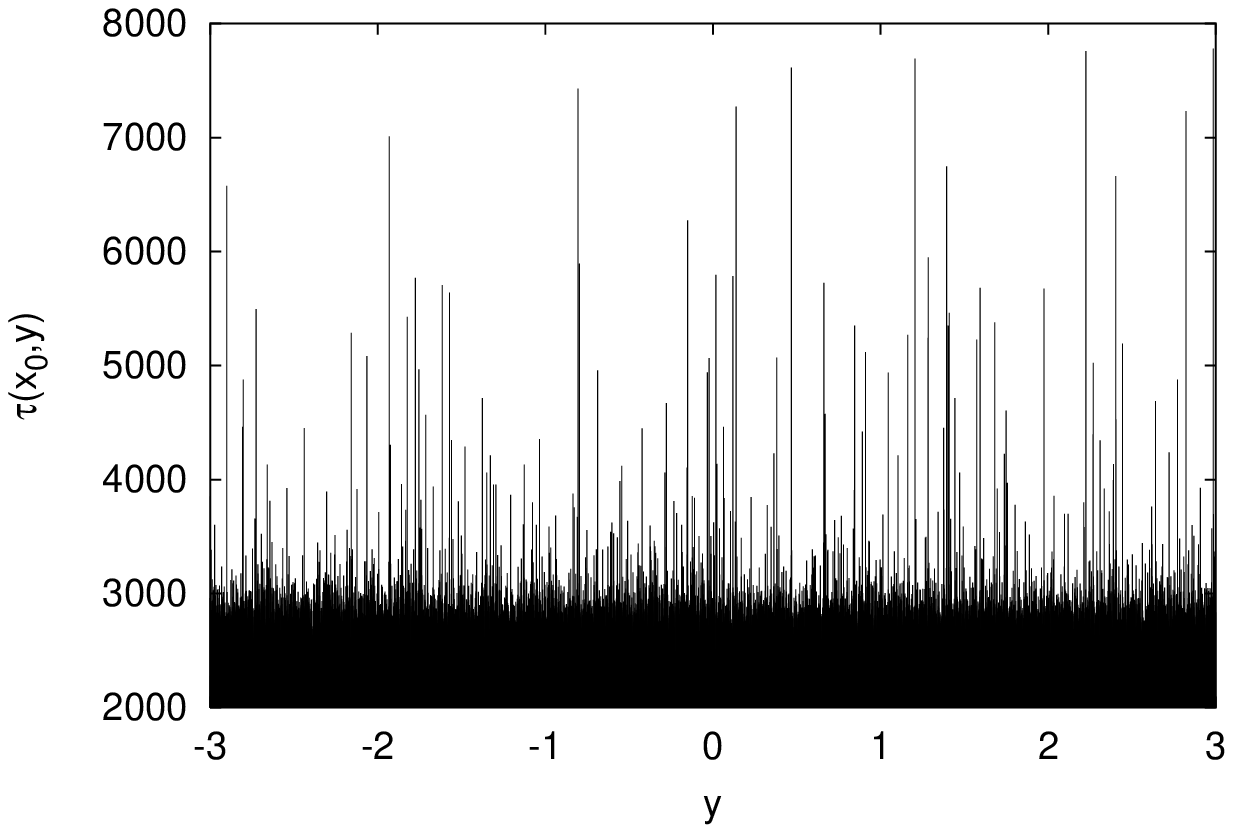,width=8cm,angle=0}}
\caption{}  
\label{fig:3}
\end{figure}

\newpage

\begin{figure}[p]

\centerline{\epsfig{figure=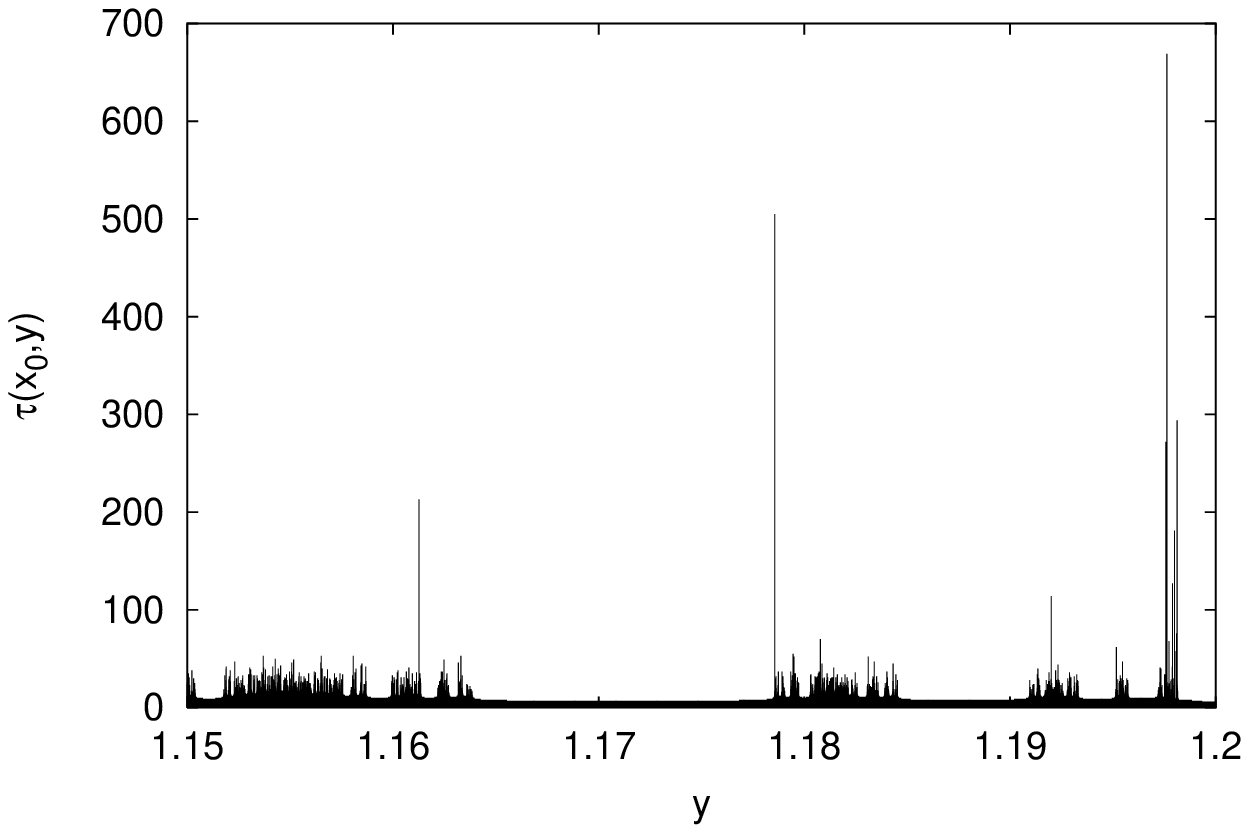,width=9cm,angle=0}}
\centerline{\hspace{.5cm} (a)}
\vspace{1.cm}
\centerline{\epsfig{figure=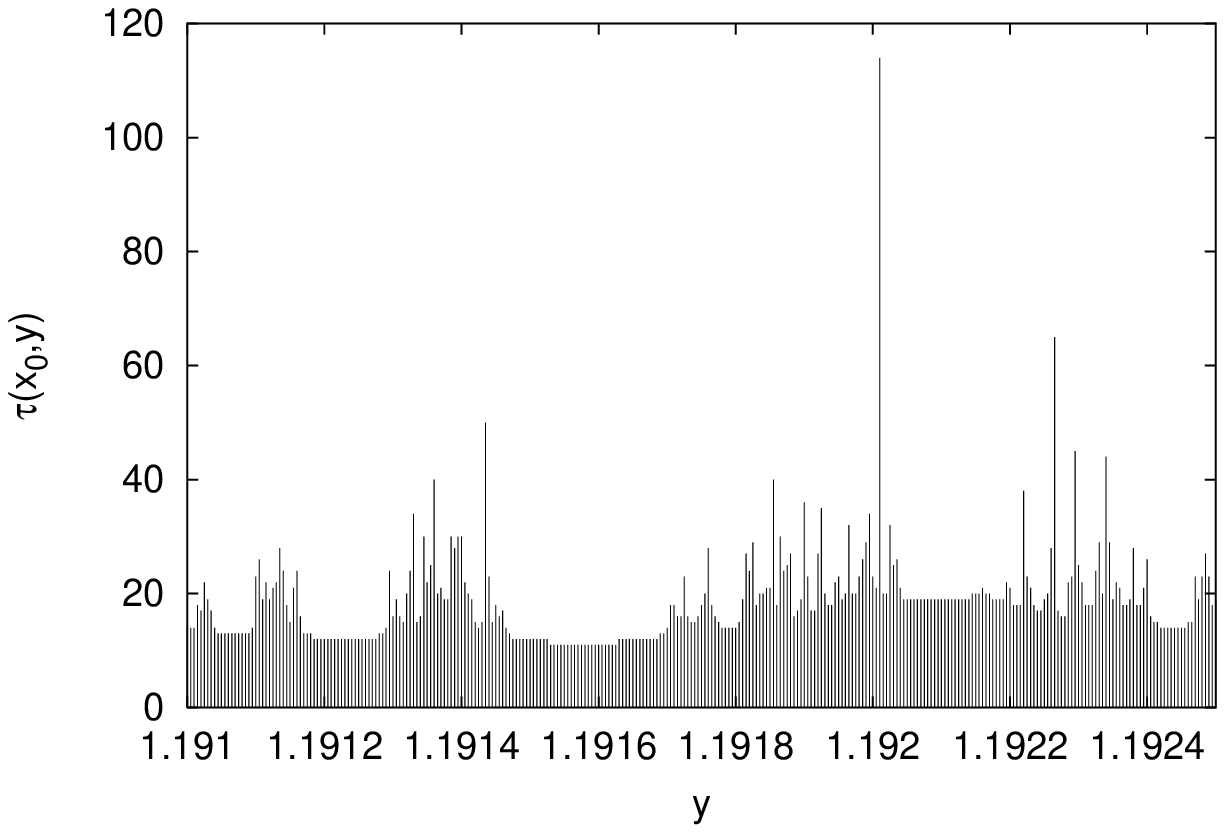,width=9cm,angle=0}}
\centerline{\hspace{.5cm} (b)}
\caption{}
\label{fig:4}
\end{figure}

\newpage 

\begin{figure}[p]
\centerline{\epsfig{figure=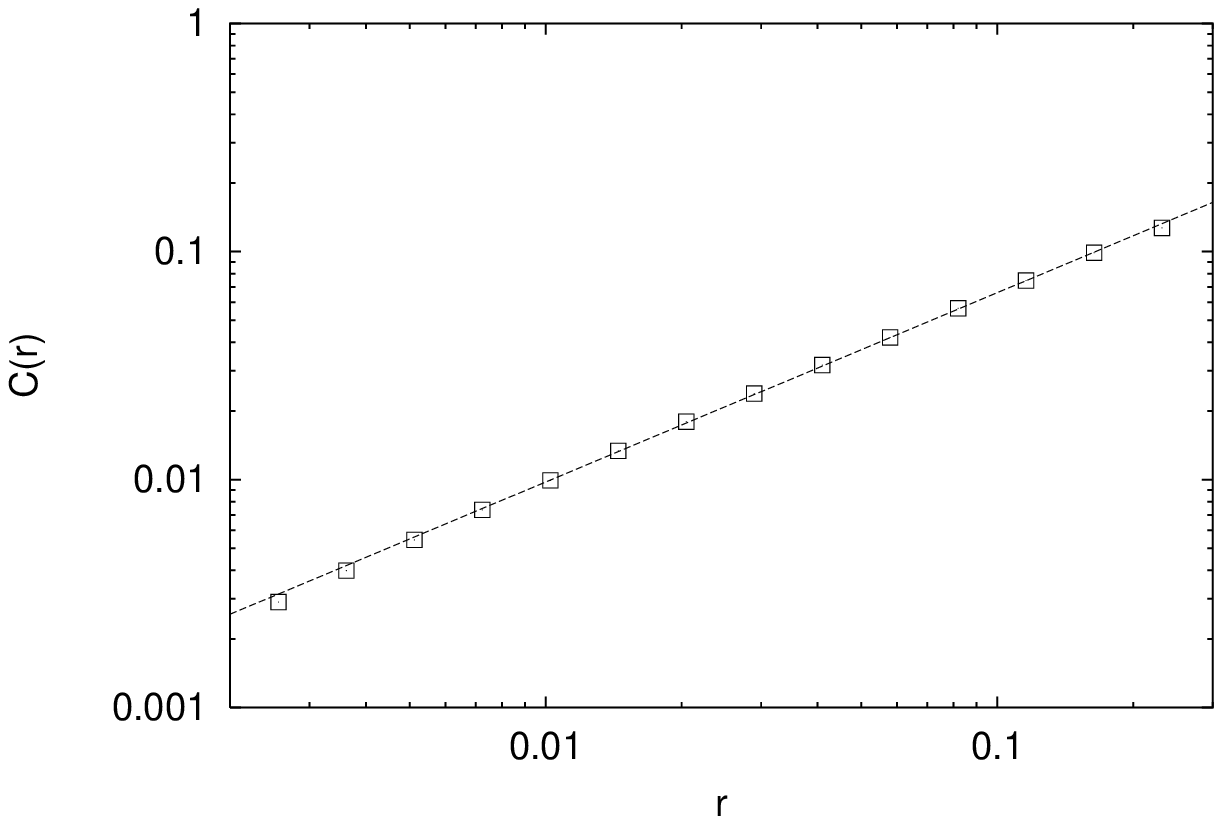,width=9cm,angle=0}}
\caption{}
\label{fig:5}
\end{figure}

\newpage 

\begin{figure}[p]
\centerline{\hspace{1cm}(a) \hspace{7.4cm}  (b)}
\centerline{\hfill  \epsfig{figure=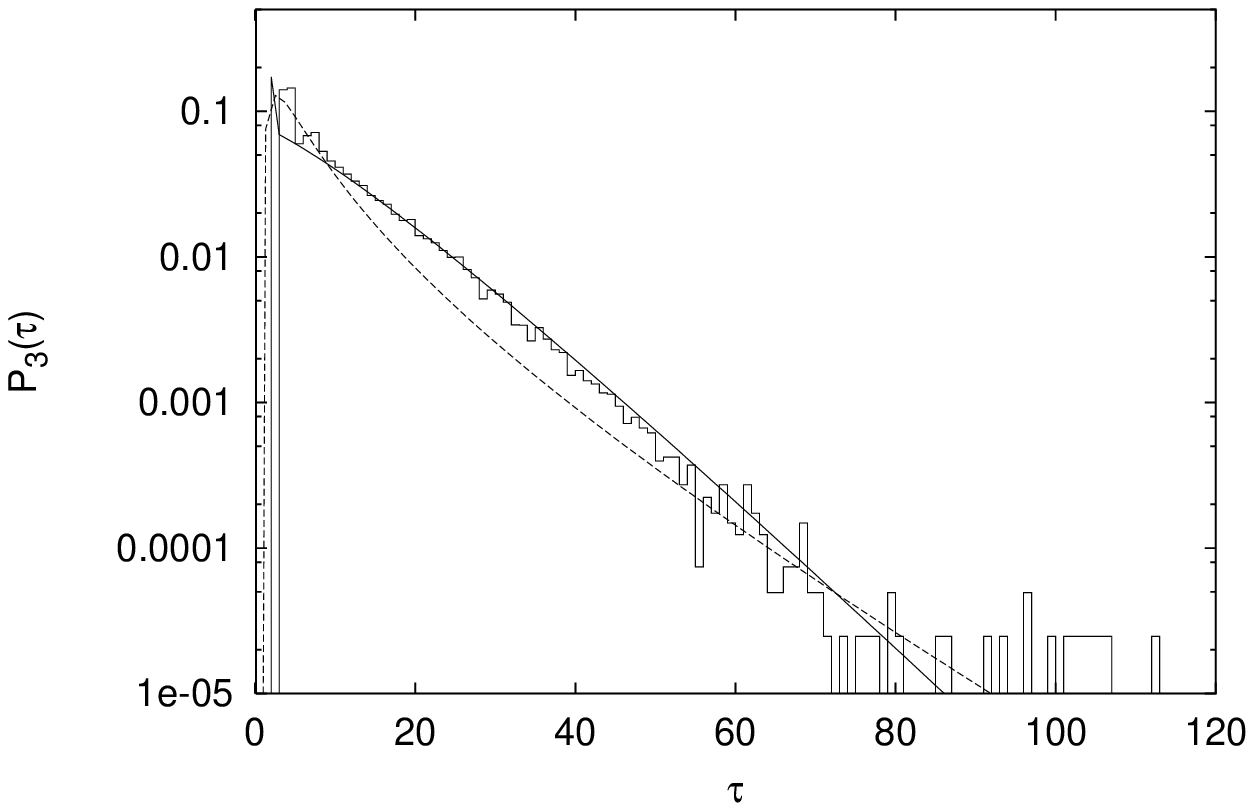,width=8cm,angle=0} \hfill 
\epsfig{figure=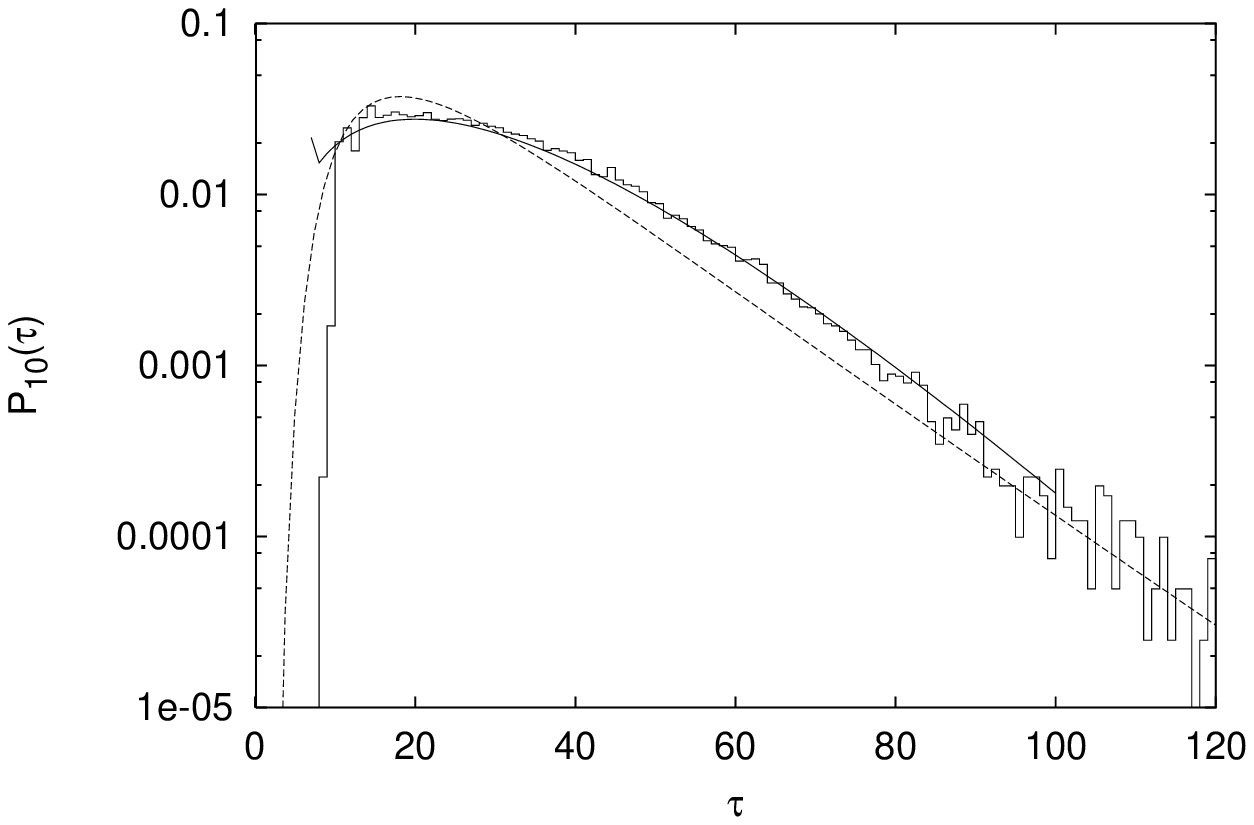,width=8cm,angle=0}}
\vspace{1.cm}
\centerline{\hspace{1cm} (c) \hspace{7.4cm}  (d)}
\centerline{\epsfig{figure=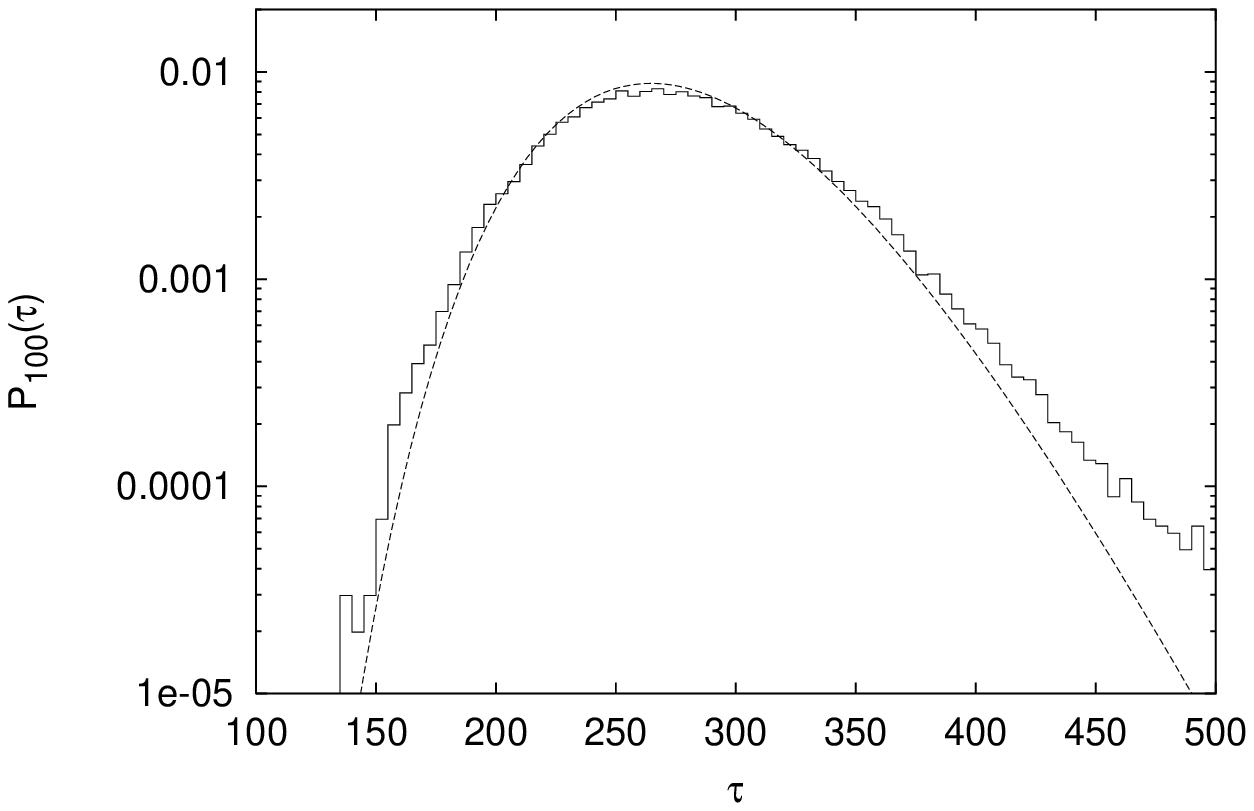,width=8cm,angle=0} \hfill 
\epsfig{figure=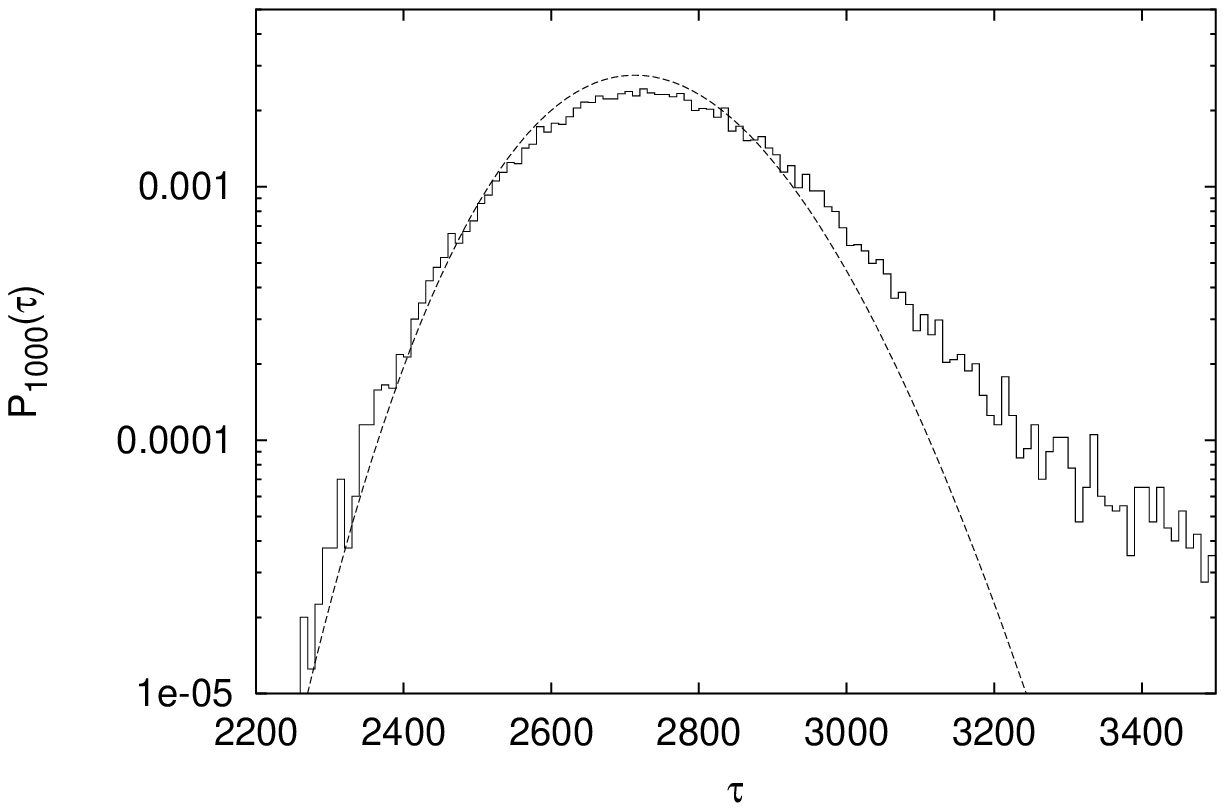,width=8cm,angle=0}}
\caption{}
\label{fig:6}
\end{figure}

\end{document}